\documentclass[aps,pra,twocolumn]{revtex4}
\newcommand {\be}{\begin{equation}}
\newcommand {\ee}{\end{equation}}
\newcommand {\bey}{\begin{eqnarray}}
\newcommand {\eey}{\end{eqnarray}}

\usepackage{epsfig}
\usepackage{amsfonts}
\usepackage{amsmath}
\usepackage{mathbbol}
\usepackage{enumerate}
\usepackage{amsthm}
\usepackage{pgfplots}
\usepackage{tikz}

\usepackage{hyperref}

\newtheorem{definition}{Definition}
\newtheorem{theorem}{Theorem}
\theoremstyle{definition}
\newtheorem{problem}{Problem}
\newtheorem{lemma}{Lemma}

\begin{document}

\title{Simple algorithm for computing the communication complexity of quantum communication processes}
\author{A. Hansen, A. Montina, S. Wolf}
\affiliation{Facolt\`a di Informatica, 
Universit\`a della Svizzera Italiana, Via G. Buffi 13, 6900 Lugano, Switzerland}

\date{\today}

\begin{abstract}
A two-party quantum communication process with classical inputs and outcomes can be 
simulated by replacing the quantum channel with a classical one. The minimal amount of 
classical communication required to reproduce the statistics of the quantum process is 
called its communication complexity. In the case of many instances simulated in parallel, 
the minimal communication cost per instance is called the asymptotic communication 
complexity. Previously, we reduced the computation of the asymptotic communication 
complexity to a convex minimization problem. In most cases, the objective function does 
not have an explicit analytic form, as the function is defined as the maximum over 
an infinite set of convex functions. Therefore, the overall problem takes the form of a 
minimax problem and cannot directly be solved by standard optimization methods. 
In this paper, we introduce a simple algorithm to compute the asymptotic communication 
complexity. For some special cases with an analytic objective function one can employ 
available convex-optimization libraries. In the tested cases our method turned out to 
be notably faster. Finally, using our method we obtain $1.238$ bits as a lower bound on 
the asymptotic communication complexity of a noiseless quantum channel with the capacity 
of 1 qubit. This improves the previous bound of $1.208$ bits.

\end{abstract}

\maketitle

\section{Introduction}
Quantum communication can be tremendously more powerful than its classical 
counterparts in solving distributed computational problems~\cite{buhrman}.
This is one of the important results of quantum
communication complexity concerned with the understanding of quantum
channels and how they compare to classical ones. 
A measure of performance of a quantum communication process --- called
the \emph{communication complexity} --- is the amount of communication
required by the most efficient classical protocol simulating the process.
If there are $N$ instances of the same quantum process, they can be 
simulated in parallel. The minimal communication cost per instance in 
the asymptotic limit $N\rightarrow \infty$ is called asymptotic 
communication complexity of the quantum process. In a previous 
work~\cite{montina}, we showed that the computation of this quantity 
can be reduced to a convex minimization problem. Generally, the objective 
function does not take an analytic form, but is given
as the maximum over an infinite set of convex functions.
Thus, the computation of the asymptotic communication complexity is
generally a minimax problem. In special cases with
a suitable symmetry, the objective function is analytically known and
the dual form of the minimization is a geometric 
program~\cite{montina2,montina3}. Geometric program is an extensively
studied class of nonlinear optimization problems and can be solved by robust 
and very efficient algorithms~\cite{gp1,gp2}. A commercial implementation is 
provided by the MOSEK package (see http://www.mosek.com). In this paper,
we present a simple and robust algorithm that solves the general 
minimax problem, which cannot be directly handled by convex
or geometric-program libraries. Furthermore, in the numerically considered 
cases in which the objective function is known, our tailored algorithm turns out 
to be much faster than available libraries solving convex problems and, more 
specifically, geometric programs.

The paper is organized as follows. In Sec.~\ref{sec_class_simu}, we introduce 
the concept of a classical simulation of a quantum communication
process and use this to define the communication complexity of
the process. In Sec.~\ref{sec_convex_opt}, we revise the results of
Ref.~\cite{montina}, where it was showed that the computation of the
communication complexity can be reduced to a convex optimization problem.
In Sec.~\ref{sec_num_algo}, the algorithm for computing the communication
complexity is presented. The convergence of the algorithm is discussed in
Sec.~\ref{convergence_section}. As the algorithm is iterative, the
solution is approached asymptotically. The iteration is stopped when
a desired accuracy is reached. In Sec.~\ref{sec_error}, we provide an
upper bound on the error and, thus, a stopping criterion.
Finally, in Sec.~\ref{num_simu_section}, we illustrate the method
with a numerical example and introduce the improved lower bound
of $1.238$ bits on the communication complexity of a noiseless quantum
channel with capacity $1$ qubit.

\section{Classical simulation of a quantum communication process}
\label{sec_class_simu}
\subsection{Quantum scenario}

Let us consider the following one-way quantum communication process between
two parties. A party, say Alice, prepares a quantum state, $a$, chosen among 
the elements of a set $A$. Then, she sends the quantum state to Bob through 
a quantum channel. Finally, Bob performs a measurement, say $b$, chosen 
among a set $B$ of measurements. The measurement produces an outcome, $s\in S$, 
with some probability $P(s|a,b)$ depending on the prepared quantum state 
and the performed measurement. The function $P(s|a,b)$ completely characterizes 
the overall process and depends on the sets $A$ and $B$ as well as the 
quantum channel used in the communication. Alice and Bob both know the sets $A$ 
and $B$, but not the choice made by the other party. That is, their choices
are not mutually conditioned.
There is no particular constraint on $A$ and $B$, but since we are interested in
implementing a numerical method, we assume that $A$ and $B$ have a finite but 
arbitrarily large number of elements. That is, their cardinalities $|A|$ and 
$|B|$ are finite. 

\subsection{Single-shot classical simulation}

Since the inputs and outcomes are classical, the statistics of 
a quantum process $P(s|a,b)$ can be classically simulated by replacing the 
quantum communication with a classical channel. Besides, Alice and 
Bob are allowed to share a stochastic random variable, say $y$. The random 
variable can be an arbitrarily long list of numbers that is generated and 
delivered to the parties before the inputs $a$ and $b$ are chosen, so that 
the numbers in the list do not contain any information on 
$a$ and $b$. The corresponding classical protocol is as follows.
Alice generates some variable $k$ according to a probability distribution
$\rho_A(k|a,y)$ that depends on Alice's input and the shared stochastic 
variable $y$. The variable $y$ is generated according to some probability 
distribution $\rho_s(y)$ and, as mentioned before, is uncorrelated 
with $a$ and $b$. Then, Alice sends $k$ to Bob. Finally, Bob generates
an outcome $s$ with a conditional probability $\rho_B(s|b,k,y)$ depending
on his input $b$, the communicated variable $k$, and the shared stochastic
variable $y$. Alice and Bob can also use private stochastic variables,
but they can be included in $y$ without loss of generality.
The protocol exactly simulates the process $P(s|a,b)$ if 
\begin{equation}
\sum_k \int dy \ \rho_B(s|b,k,y)\rho_A(k|a,y)\rho_s(y)=
P(s|a,b).
\end{equation}
Note that the integral symbol stands for integral over some measurable space, 
but the space of $y$ could be indifferently discrete.

There are different definitions of communication cost of a protocol. Without
loss of generality, we can assume that $k$ is generated deterministically
for each value of Alice's input $a$ and random variable $y$. 
The number of bits required to encode and transmit the variable $k$ depends 
in general on these two values.
Let $C(a,y)$ be the number of bits sent by Alice when she chooses
$a$ with the shared noise $y$. The {\it worst-case communication cost} is the 
maximum of $C(a,y)$ over every possible value taken by $y$ and $a$.
Alternatively, one can first average over $y$ and then take the maximum
over the input $a$ to obtain the so-called {\it average communication cost}
\begin{equation}
\bar {\cal C}\equiv \max_{a} \int dy \rho_s(y) C(a,y).
\end{equation}
There is also an entropic definition, which has been used in Ref.~\cite{montina}. 
The entropic cost is always smaller than or equal to the average cost
$\bar{\cal C}$. The results which are presented here hold for both of the last
two definitions, thus the average and entropic costs can be used indifferently.
Here, we will refer to the average cost $\bar {\cal C}$.
\begin{definition}[Communication complexity]
We define the communication complexity ${\cal C}_{min}$ of a quantum process
$P(s|a,b)$ to be the minimal communication cost required to simulate it.
\end{definition}

\subsection{Parallel protocols}

If the two parties simulate $N$ instances of the same quantum process
$P(s|a,b)$ with $N$ different inputs $a$ and $b$ for each instance,
it is possible to envisage a larger set of communication protocols, where the 
probability of generating $k$ can depend on the full set of Alice's inputs,
$a^{i=1,2,\dots,N}$. In other words, the distribution $\rho_A(k|a,y)$ becomes
$\rho_A(k|a^1,a^2,\dots,a^{N},y)$. The asymptotic communication cost, hereafter
denoted by ${\cal C}^{asym}$, is the cost of the parallelized simulation divided by 
$N$ in the limit of $N\rightarrow\infty$.
\begin{definition}
We define the asymptotic communication complexity ${\cal C}^{asym}_{min}$ of a
problem $P(s|a,b)$ to be the minimum of ${\cal C}^{asym}$ over the class of parallel 
protocols that solve the problem.
\end{definition}
Since the set of protocols working for parallel simulations is larger than the set of 
single-shot protocols, it is clear that
\begin{equation}
{\cal C}_{min}^{asym}\le {\cal C}_{min}.
\end{equation}
However, as showed in Ref.~\cite{montina}, the difference between ${\cal C}_{min}^{asym}$ 
and ${\cal C}_{min}$ scales at most as the logarithm of ${\cal C}_{min}^{asym}$,
as revised in the next section.

\section{Computation of ${\cal C}_{min}^{asym}$ as a convex optimization problem}
\label{sec_convex_opt}

In Ref.~\cite{montina}, we showed that the computation of the asymptotic
communication complexity ${\cal C}_{min}^{asym}$ is equivalent to a convex 
optimization problem. Tight lower and upper bounds on the single-shot communication 
complexity ${\cal C}_{min}$ are given in terms of ${\cal C}_{min}^{asym}$.
The optimization is made over a suitable set of probability distributions. The set, 
denoted by ${\cal V}(P)$, depends on the quantum process $P$ and is defined as 
follows.
\begin{definition}
Given a process $P(s|a,b)$, the set ${\cal V}(P)$ is defined as the set of
conditional probabilities $\rho({\bf s}|a)$ over the sequence 
${\bf s}=\{s_1,\dots,s_{|B|}\}\in S^{|B|}$
whose marginal distribution of the $b$-th variable is equal to $P(s|a,b)$.
In other words, the set ${\cal V}(P)$ contains any $\rho({\bf s}|a)$ satisfying the 
constraints
\begin{equation}
\label{constraints}
\sum_{{\bf s},s_b=s} \rho({\bf s}|a)=P(s|a,b)\quad \forall a,b \text{ and } s,
\end{equation}
where the sum is performed over every element of the sequence $\bf s$ except the
$b$-th element $s_b$, which is set equal to $s$.
\end{definition}

The central result in Ref.~\cite{montina} is a convex optimization problem that yields the asymptotic communication 
complexity of $P$. The asymptotic communication complexity is equal to the minimum of the capacity of the channels 
$\rho({\bf s}|a)\in {\cal V}(P)$ --- a convex functional over 
${\cal V}(P)$. Before introducing the definition of the channel
capacity, let us recall some concepts from information theory. The mutual 
information of two stochastic variables $X$ and $Y$ is defined as
\begin{equation}
I(X;Y)=\sum_x\sum_y \rho(x,y)\log_2\frac{\rho(x,y)}{\rho(x)\rho(y)},
\end{equation}
where $\rho(x,y)$ is the joint probability distribution of $x$ and $y$,
and $\rho(x)$ and $\rho(y)$ are the marginal distributions of $x$ and $y$,
respectively~\cite{cover}. The mutual information is a measure of the degree of
correlation between two stochastic variables. It is always non-negative, and
equal to zero if the variables are uncorrelated. Let us now introduce
the concept of a channel. In information theory, a channel is a physical 
device, such as a wire, carrying information from a sender to a receiver. 
The channel is mathematically represented by a conditional probability
$\rho(y|x)$ of getting the outcome $y$ given the input $x$~\cite{cover}. 
The capacity of the channel $x\rightarrow y$, which we denote by
$C(x\rightarrow y)$, is the maximum of the mutual information
between $x$ and $y$ over the space of probability distributions
$\rho(x)$ of the input $x$~\cite{cover}, that is,
\begin{equation}
\label{capacity}
C(x\rightarrow y)\equiv \max_{\rho(x)} I(x;y).
\end{equation}
The information-theoretic interpretation of the channel capacity is provided 
by the noisy-channel coding theorem~\cite{cover}. Roughly speaking, the 
capacity of a channel is the maximum rate of information that can be 
transmitted through the channel. 

Given these definitions, let us introduce the functional ${\cal D}(P)$
as the minimum of the capacity over the distributions 
$\rho({\bf s}|a)\in {\cal V}(P)$.
\begin{eqnarray}
\label{cal_D}
{\cal D}(P)\equiv\min_{\rho({\bf s}|a)\in {\cal V}(P)}
C(a\rightarrow {\bf s})=  \\
\nonumber
\min_{\rho({\bf s}|a)\in {\cal V}(P)}\max_{\rho(a)} I(A;{\bf S}).
\end{eqnarray}

The following theorems, proved in Ref.~\cite{montina}, relate ${\cal D}(P)$
to the communication complexity.
\begin{theorem}
\label{main_theor}
The asymptotic communication complexity ${\cal C}_{min}^{asym}$ of $P$ is equal 
to ${\cal D}(P)$.
\end{theorem}
\begin{theorem}
The communication complexity  ${\cal C}_{min}$ is bounded by the inequalities
\begin{equation}
{\cal D}(P)\le {\cal C}_{min}\le {\cal D}(P)+2\log_2[{\cal D}(P)+1]+2\log_2 e.
\end{equation}
\end{theorem}
The single-shot communication complexity ${\cal C}_{min}$ is always greater than or 
equal to the asymptotic communication complexity ${\cal C}_{min}^{asym}$. However,
as anticipated in the previous section, the difference scales at most logarithmically
in ${\cal C}_{min}^{asym}$. 
Theorem~\ref{main_theor} reduces the computation of the asymptotic communication
complexity to the following convex optimization problem.
\begin{problem}
\label{prob1}
\begin{equation}
\begin{array}{c}
\min_{\rho({\bf s}|a)} C(a\rightarrow{\bf s})  \\
\text{subject to the constraints}  \\
\rho({\bf s}|a)\ge0,  \\
\sum_{{\bf s},s_b=s} \rho({\bf s}|a)=P(s|a,b).
\end{array}
\end{equation}
\end{problem}
Note that the functional $C(a\rightarrow {\bf s})$ is convex in $\rho({\bf s}|a)$,
since the mutual information is convex in $\rho({\bf s}|a)$~\cite{cover} and
the pointwise maximum of a set of convex functions is a convex 
function~\cite{boyd}.

In general, the channel capacity does not have a known analytic expression. 
However, in some symmetric problems, it is possible to get rid of the 
maximization over $\rho(a)$ in the definition of the channel capacity given by 
Eq.~(\ref{capacity}). This can be shown by using Sion's minimax theorem~\cite{minimax} 
and some general properties of the mutual information.
As the mutual information is convex in $\rho({\bf s}|a)$ and concave in 
$\rho(a)$~\cite{cover}, 
we have from the minimax theorem that the minimization and maximization 
in the definition of ${\cal D}(P)$ [Eq.~(\ref{cal_D})] can be interchanged. Thus, 
we obtain
\begin{equation}
\label{swap_minmax}
{\cal D}(P)=\max_{\rho(a)} {\cal J}(P)
\end{equation}
where 
\begin{equation}
\label{eq_J}
{\cal J}(P)\equiv  \min_{\rho({\bf s}|a)\in{\cal V}(P)} I(A;{\bf S})
\end{equation}
is a functional of $\rho(a)$. As $I(A;{\bf S})$ is concave in $\rho(a)$ and the pointwise 
minimum of a set of concave functions is concave~\cite{boyd}, the functional ${\cal J}(P)$ 
is concave.
In some cases, it is trivial to find the distribution $\rho_{max}(a)$ maximizing ${\cal J}(P)$. 
For example, if the conditional probability $P(s|a,b)$ is invariant under the transformation
$a\rightarrow a+1$ up to some suitable transformation of $b$ and $s$, then 
we can infer by symmetry and the concavity of ${\cal J}(P)$ that the uniform 
distribution maximizes ${\cal J}(P)$. This case will be considered as a numerical
example in Sec.~\ref{num_simu_section}.

Thus, if $\rho_{max}(a)$ is known, the computation of ${\cal C}_{min}^{asym}$ is reduced 
to the following convex optimization problem.
\begin{problem}
\label{prob2}
\begin{equation}
\begin{array}{c}
\min_{\rho({\bf s}|a)} I(A;{\bf S})  \\
\text{subject to the constraints}  \\
\rho({\bf s}|a)\ge0,  \\
\sum_{{\bf s},s_b=s} \rho({\bf s}|a)=P(s|a,b).
\end{array}
\end{equation}
\end{problem}

As shown in Refs.~\cite{montina2,montina3}, the dual form of Problem~\ref{prob2}
is a geometric program (see Ref.~\cite{boyd}
for an introduction to dual theory). Geometric program is an extensively
studied class of nonlinear optimization problems~\cite{gp1,gp2} and the
commercial package MOSEK (see http://www.mosek.com) provides a solver
specialized for this class. However, if the distribution $\rho_{max}(a)$
is not known and we set $\rho(a)$ equal to an arbitrary distribution,
the solution of Problem~2 yields merely a lower bound on
the asymptotic communication complexity.

In Sec.~\ref{sec_num_algo}, we present a simple and robust algorithm that \emph{directly}
solves Problem~\ref{prob1}. Furthermore, in Sec.~\ref{num_simu_section}, we
consider some numerical situations in which $\rho_{max}(a)$ is known, and
we show that the introduced algorithm turns out to be much faster than 
the Mosek package in solving Problem~\ref{prob2}.

\section{Numerical algorithm}
\label{sec_num_algo}

We introduce a simple numerical algorithm (herafter Algorithm~1) for solving 
Problem~\ref{prob1} and computing the asymptotic communication complexity of a quantum 
process $P(s|a,b)$. A further simplification is given by Algorithm~2, which solves
Problem~\ref{prob2}. The two algorithms are based 
on the \emph{block coordinate descent method}~\cite{bertsekas}, also called alternating
minimization, block-nonlinear Gauss-Seidel method or block coordinate descent method. 
The alternating minimization is an iterative method that performs the minimization over 
blocks of variables. Namely, the set of variables, with respect to which the minimization 
is performed, is divided in blocks, say $X_1, X_2,\dots X_W$.
The objective function is first minimized with respect to the variables in the block $X_1$, while 
keeping the variables in the other blocks constant, then with respect to the variables in $X_2$ and 
so on and so forth. This procedure is repeated cyclically.
There are several
results on the convergence of the method for constrained and unconstrained problems.
The continuous differentiability of the objective function is generally the basic common 
assumption. In Ref.~\cite{bertsekas}, it is proved that the algorithm converges toward
a minimum if each block minimization has a unique solution. As the objective function 
of our problem is not differentiable everywhere (see later discussion), the results relying 
on the continuous differentiability cannot be employed for a convergence proof. The 
convergence of Algorithm~2 is a consequence of a general theorem proved in Ref.~\cite{csiszar}.
Although these results cannot be adapted to the case of Algorithm~1,
we will provide arguments for the convergence in
Sec.~\ref{convergence_section}. The convergence proof for Algorithm~2 is given in the
same section.

The alternating minimization method with a two-block partition is particularly advantageous
for the computation of the asymptotic communication complexity. Indeed, using a suitable
partition, one block minimization turns out to be decoupled from the maximization with
respect to $\rho(a)$, whereas the minimization in the other block can be performed 
analytically. 

To derive the algorithm, let us recast Problem~\ref{prob1} as follows. The task is
to evaluate the quantity ${\cal D}(P)$ defined by Eq.~(\ref{cal_D}), which takes the
form of Eqs.~(\ref{swap_minmax},\ref{eq_J}).
The mutual information $I(A;{\bf S})$ can be rewritten as minimization of the functional
\begin{equation}
\label{new_obj_funct}
{\cal K}\equiv
\sum_{{\bf s},a}\rho({\bf s}|a)\rho(a)\ln\frac{\rho({\bf s}|a)}{R({\bf s})}
\end{equation}
over the space of probability distributions $R({\bf s})$. Indeed, using the
Karush-Kuhn-Tucker conditions for optimality~\cite{boyd}, we find that 
the global minimum of the functional $\cal K$ with respect to $R({\bf s})$ is at 
$R({\bf s})=\rho({\bf s})$ [note that the functional is convex
in $R({\bf s})$]. Thus, Eqs.~(\ref{swap_minmax},\ref{eq_J}) turn into the
following minimax problem,
\begin{equation}
{\cal D}(P)=\max_{\rho(a)}\min_{\rho({\bf s}|a)\in{\cal V}(P)}\min_{R({\bf s})} {\cal K}.
\end{equation}
As $\cal K$ is linear in $\rho(a)$ and convex in $\rho({\bf s}|a)$ and $R({\bf s})$,
we can swap again the minimization and the maximization~\cite{minimax}, and obtain
\begin{equation}
\label{min_K0}
{\cal D}(P)=\min_{\rho({\bf s}|a)\in{\cal V}(P)}\min_{R({\bf s})} \bar{\cal K},
\end{equation}
where 
\begin{equation}
\label{funct_I0}
\bar{\cal K}\equiv \max_{\rho(a)} {\cal K}
\end{equation}
is a convex functional of $\rho({\bf s}|a)$ and $R({\bf s})$. 
Note that the function $\bar{\cal K}$
is not differentiable if $\rho({\bf s}|a)$ or $R({\bf s})$ are zero for some $\bf s$ and $a$.
Furthermore, $\bar{\cal K}$ is not differentiable in other points, since the maximizing
distribution $\rho(a)$ can be discontinuous as a function of $\rho({\bf s}|a)$ and
$R({\bf s})$. 

To find the global minimum of $\bar{\cal K}$, we apply the block coordinate descent method 
by alternately minimizing with respect to $\rho({\bf s}|a)$ and $R({\bf s})$. Given 
a strictly positive initial distribution $R({\bf s})$ (the strict positivity 
is fundamental for the convergence, as discussed in the end of the section
and in Sec.~\ref{convergence_section}),
we search for the distribution $\rho({\bf s}|a)$ minimizing $\bar{\cal K}$ 
over the set ${\cal V}(P)$. Then, we minimize with respect to $R({\bf s})$. We iterate
by using these two minimization steps until we get the global minimum up to some given 
accuracy. By construction, each iteration always lowers the value of $\bar{\cal K}$.

\subsection{Minimization w.r.t $\rho({\bf s}|a)$}
\label{min_rho_sa}
Let us consider the minimization with respect to $\rho({\bf s}|a)$. This is made
in the domain of non-negative functions under the constraints of Problem~1
and corresponds to solve the minimax problem
\begin{equation}
\label{minimax_algo}
\min_{{\rho({\bf s}|a)}}\max_{\rho(a)} {\cal K}=
\max_{\rho(a)}\min_{{\rho({\bf s}|a)}}{\cal K}.
\end{equation}
We first solve the minimization problem, and show that the solution does
not depend on the distribution $\rho(a)$. The distribution $\rho({\bf s}|a)$
solving the minimization problem under the constraints~(\ref{constraints})
minimizes the Lagrangian 
\begin{equation}
\label{lagr_1}
{\cal L}_1\equiv {\cal K}-\sum_{s,a,b}\bar\lambda(s,a,b)\rho(a)\left[\sum_{{\bf s},s_b=s}
\rho({\bf s}|a)-P(s|a,b)\right] 
\end{equation}
over the domain of $\cal K$,
where $\bar\lambda$ are suitable Lagrange multipliers that are set 
so that the constraints~(\ref{constraints}) are satisfied or, equivalently,
by maximizing the dual objective function, as discussed later.
To find the minimum, we set the derivative of ${\cal L}_1$ with 
respect to $\rho({\bf s}|a)$ equal to zero and obtain an optimal 
distribution of the form
\begin{equation}
\label{min_1}
\rho({\bf s}|a)=R({\bf s})e^{\sum_b\lambda(s_b,a,b)},
\end{equation}
where $\lambda\equiv\bar\lambda-1/|B|$. Replacing the distribution in
${\cal L}_1$, we obtain
\begin{equation}
\label{Eq_K_1}
{\cal K}_1\equiv\sum_{s,a,b}P(s|a,b)\rho(a)\lambda(s,a,b)+1-\sum_{\bf s}R({\bf s}) 
F_\lambda({\bf s}),
\end{equation}
where
\begin{equation}
F_\lambda({\bf s})\equiv \sum_a\rho(a)e^{\sum_b\lambda(s,a,b)}.
\end{equation}
By definition, ${\cal K}_1$ is the dual objective function of the original
minimization problem (see Ref.~\cite{boyd} for an introduction to dual theory).
Since the constraints of the primal problem satisfy
the Slater conditions~\cite{boyd}, strong duality holds,
and the maximum of ${\cal K}_1$ is equal to the minimum in the original 
problem. The maximum is characterized by setting the derivative
with respect to $\lambda$ equal to zero. This gives 
\begin{equation}
\label{eq_constr_algorithm}
\sum_{{\bf s},s_b=s} R({\bf s})e^{\sum_b\lambda(s_b,a,b)}=
P(s|a,b),
\end{equation}
that is, the constraints~(\ref{constraints}), as shown by Eq.~(\ref{min_1}).
Thus, the minimizing distribution $\rho({\bf s}|a)$ is given by Eq.~(\ref{min_1})
and the solution of Eq.~(\ref{eq_constr_algorithm}).
As ${\cal K}_1$ is a concave function, solving Eq.~(\ref{eq_constr_algorithm}) 
is equivalent to an unconstrained convex optimization problem, and it can be 
solved through the Newton method~\cite{boyd}. The introduced quantity 
$F_\lambda({\bf s})$ plays an important role in the dual form of Problem~2, 
in evaluating lower and upper bounds on the asymptotic communication 
complexity and in the formulation of necessary and sufficient conditions 
for optimality.

At this point, it is important to stress that the solution of 
Eq.~(\ref{eq_constr_algorithm}) does not depend on $\rho(a)$. That is,
the minimization of $\cal K$ in the minimax problem~(\ref{minimax_algo})
is completely decoupled from the maximization over $\rho(a)$. Thus,
we have reduced the first step of the algorithm to a simple unconstrained
maximization of the function ${\cal K}_1$ with respect to $\lambda(s,a,b)$.
The computation of the optimal $\rho(a)$ is irrelevant, as it does not
affect the next step, the minimization with respect to $R({\bf s})$.

\subsection{Minimization w.r.t. $R({\bf s})$}
Let us consider the minimization of $\bar{\cal K}$ with respect to 
$R({\bf s})$, which corresponds to solve the minimax problem
\begin{equation}
\label{minimax_algo_2}
\min_{R({\bf s})}\max_{\rho(a)} {\cal K}=
\max_{\rho(a)}\min_{R({\bf s})}{\cal K}.
\end{equation}
As we already said, the minimization with respect to $R({\bf s})$ yields
\begin{equation}
R({\bf s})=\sum_a\rho({\bf s}|a)\rho(a)\equiv\rho({\bf s}).
\end{equation}
Thus, the minimization replaces $R({\bf s})$ with $\rho({\bf s})$. Making this
replacement in ${\cal K}$, we get from Eq.~(\ref{new_obj_funct}) 
\begin{equation}
{\cal K}=I({\bf S};A)=
\sum_{{\bf s},a}\rho({\bf s}|a)\rho(a)\ln\frac{\rho({\bf s}|a)}{\rho({\bf s})},
\end{equation}
that is, ${\cal K}$ is the mutual information between the variables $a$ and 
$\vec s$. Thus, the maximization with respect to $\rho(a)$ in 
Eq.~(\ref{minimax_algo_2}) is just the computation of the capacity of the
channel $\rho({\vec s}|a)$, which can be performed by using standard methods,
such as the Blahut-Arimoto algorithm~\cite{cover}. 

\subsection{Alternating minimization}
The two block-minimizations over $\rho({\bf s}|a)$ and $R({\bf s})$ 
are iterated until a given accuracy is reached. The stopping criteria will be
discussed in Sec.~\ref{sec_error}.

Summarizing, the algorithm is as follows. \newline
{\bf Algorithm 1} (for Problem~\ref{prob1}). \newline
\begin{enumerate}
\item Set some initial distribution $R({\bf s})>0$.
\item Compute $\lambda(s,a,b)$ solving the equations
\begin{equation}
\sum_{{\bf s},s_b=s} R({\bf s})e^{\sum_{\bar b}\lambda(s_{\bar b},a,\bar b)}=P(s|a,b).
\end{equation}
\item\label{step_rho_sa} Set $\rho({\bf s}|a)=R({\bf s}) e^{\sum_b\lambda(s_b,a,b)}$.
\item 
\label{maxi_step}
Maximize the mutual information
\begin{equation}
\label{mutual_algo}
I({\bf S};A)=\sum_{{\bf s},a} \rho({\bf s}|a)\rho(a)\log
\frac{\rho({\bf s}|a)}{\sum_{\bar a}\rho({\bf s}|\bar a)\rho(\bar a)}
\end{equation}
with respect to $\rho(a)$ [computation of the capacity of the channel $\rho({\bf s}|a)$].
\item Set $R({\bf s})=\sum_a\rho({\bf s}|a)\rho(a)$.
\item\label{stop_item} Stop if a given accuracy is reached.
\item Repeat from step~2.
\end{enumerate}

The algorithm can be recast into a more illuminating form by getting rid of
$\rho({\bf s}|a)$. This also reduces the amount of required memory by about 
a factor of $|A|$. Furthermore, the new form suggests an approximate computation
of the channel capacity that is much more efficient numerically. Using 
constraint~(\ref{constraints}) and the expression of $\rho({\bf s}|a)$ given at 
step~\ref{step_rho_sa}, the mutual information~(\ref{mutual_algo}) 
takes the form
\begin{equation}
\label{exact_I}
\begin{array}{c}
I({\bf S};A)=\sum_{s,a,b}P(s|a,b)\rho(a)\lambda(s,a,b)-  \vspace{1mm} \\
\sum_{\bf s}R({\bf s}) F_\lambda({\bf s})\log F_\lambda({\bf s}).
\end{array}
\end{equation}
As shown in Refs.~\cite{montina2,montina3} and later in 
Secs.~\ref{conv_algo1} and \ref{sec_nec_suff},
the optimal solution 
the minimizer satisfies the equation $R({\bf s})\left[F_\lambda({\bf s})-1\right]=0$.
Thus, if $R({\bf s})$ and $\lambda(s,a,b)$ are close to the solution, we can
approximate $I({\bf S};A)$ up to the second order in $F_\lambda({\bf s})-1$,
\begin{equation}
\begin{array}{c}
I({\bf S};A)\simeq \sum_{s,a,b}P(s|a,b)\rho(a)\lambda(s,a,b)+  \vspace{1mm} \\
\sum_{\bf s}R({\bf s}) F_\lambda({\bf s})\left[1-F_\lambda({\bf s})\right].
\end{array}
\end{equation}
This form is quadratic in $\rho(a)$. Using Eq.~(\ref{constraints}) and the
definition of $F_\lambda({\bf s})$, we obtain
\begin{equation}
\label{quadratic_form}
I({\bf S};A)\simeq \sum_{a} d_1(a)\rho(a)+1-\sum_{a,a'}d_2(a,a')\rho(a)\rho(a'),
\end{equation}
where 
\begin{equation}
\label{quadr_form}
\begin{array}{c}
d_1(a)\equiv \sum_{s,b}P(s|a,b)\lambda(s,a,b) \vspace{1mm} \\
d_2(a,a')\equiv \sum_{\bf s}R({\bf s})e^{\sum_b\lambda(s_b,a,b)+\sum_b\lambda(s_b,a',b)}.
\end{array}
\end{equation}
To maximize the quadratic form~(\ref{quadratic_form}) is numerically much more efficient
than maximizing the exact form~(\ref{exact_I}). Indeed, the maximization of the exact
form requires to compute the objective function and, possibly, its derivatives many
times. The associated computational cost grows exponentially with $|B|$ because of the sum
over $\bf s$. Conversely, with the approximate form, computation of the coefficients $d_1$ 
and $d_2$, which is the hardest part, is made only once before each maximization.

Numerical experiments show that this approximation does not affect the
convergence. Algorithm~1 is recast as follows \newline
{\bf Algorithm 1b} (for Problem~\ref{prob1}). \newline
\begin{enumerate}
\item Set some initial distribution $R({\bf s})>0$.
\item Compute $\lambda(s,a,b)$ solving the equations
\begin{equation}
\sum_{{\bf s},s_b=s} R({\bf s})e^{\sum_{\bar b}\lambda(s_{\bar b},a,\bar b)}=P(s|a,b).
\end{equation}
\item\label{maxi_step2}
Maximize the function $I({\bf S};A)$, given by Eq.~(\ref{exact_I}) or its approximate 
form~(\ref{quadratic_form}), with respect to $\rho(a)$.
\item\label{update_R}
Perform the replacement $R({\bf s})\rightarrow R({\bf s}) F_\lambda({\bf s})$.
\item Stop if a given accuracy is reached.
\item Repeat from step~2.
\end{enumerate}
This recast of Algorithm will be useful for the subsquent discussion on the convergence.

If the distribution $\rho(a)$ is known, we can fix it and skip step~\ref{maxi_step2}
performing the maximization of $I({\bf S};A)$ over $\rho(a)$. The resulting algorithm
solves Problem~\ref{prob2} and is as follows.
\newline
{\bf Algorithm 2} (for Problem~\ref{prob2}). \newline
\begin{enumerate}
\item Set some initial distribution $R({\bf s})>0$.
\item Compute $\lambda(s,a,b)$ solving the equations
\begin{equation}
\sum_{{\bf s},s_b=s} R({\bf s})e^{\sum_{\bar b}\lambda(s_{\bar b},a,\bar b)}=P(s|a,b).
\end{equation}
\item Perform the replacement $R({\bf s})\rightarrow R({\bf s}) F_\lambda({\bf s})$.
\item Stop if a given accuracy is reached.
\item Repeat from step~2.
\end{enumerate}

It is worth to note that the initialization $R({\bf s})>0$ is necessary for the 
convergence of the algorithm, unless the domain of the minimizer distribution, say 
$\rho(\bf s)$, is known. Indeed, suppose that we set $R({\bf s}')=0$ for some ${\bf s}'$,
but $\rho({\bf s}')\ne 0$. The update performed at step~\ref{update_R}
of Algorithm~1b keeps $R({\bf s})=0$, provided that $F_\lambda({\bf s})$ is finite.
Thus, the algorithm never converges toward the solution.

\section{Convergence proof}
\label{convergence_section}
The convergence of Algorithm~2 is a consequence of the results in Ref.~\cite{csiszar}
and will be proved below. Although this proof does not hold for the
general Problem~1, in the end of the section, we will give some arguments for
the convergence of Algorithm~1.

The proof relies on three simple lemmas.
\begin{lemma}\label{lemma_csis} (Lemma~1 in Ref.~\cite{csiszar}) Let $a_n$ and $b_n$ ($n=0,1,\dots$) be
extended real numbers greater than $-\infty$ and $c$ a finite number such that
\begin{equation}
\label{recur_eq}
c+b_{n-1}\ge b_n+a_n,\;\; n=1,2,\dots
\end{equation}
and 
\begin{equation}
\limsup\limits_{n\rightarrow\infty} b_n>-\infty, \;\; b_{n_0}<+\infty \text{  for some  }n_0.
\end{equation}
Then, 
\begin{equation}
\liminf_{n\rightarrow\infty} a_n \le c.
\end{equation}
\end{lemma}
{\it Proof.} Suppose that 
$\liminf_{n\rightarrow\infty} a_n > c$. As $b_{n_0}$ is finite, Eq.~(\ref{recur_eq}) implies
that $b_n$ is finite for every $n\ge n_0$. Furthermore, from Eq.~(\ref{recur_eq}) we have that
\begin{equation}
\liminf_{n\rightarrow\infty}(b_{n-1}-b_n)>0,
\end{equation}
which implies that $\lim_{n\rightarrow\infty} b_n=-\infty$, in contradiction to the
hypothesis. $\square$\newline

\begin{lemma}
\label{lemma_1_ineq}
Let $\rho_1({\bf s}|a)$ be the minimizer of 
\begin{equation}
\left. {\cal K}\right|_{R=R_0}
\equiv \sum_{{\bf s},a}\rho({\bf s}|a)\rho(a)\log\frac{\rho({\bf s}|a)}{R_0({\bf s})}
\end{equation}
\end{lemma}
with respect to $\rho({\bf s}|a)\in{\cal P}$, where $\cal P$ is a convex set. Then,
\begin{equation}
\label{ineq_lemma2}
\sum_{{\bf s},a}\rho_1({\bf s}|a)\rho(a)\log\frac{\rho_1({\bf s}|a)}{R_0({\bf s})}\le
\sum_{{\bf s},a}\rho({\bf s}|a)\rho(a)\log\frac{\rho_1({\bf s}|a)}{R_0({\bf s})}
\end{equation}
for every $\rho({\bf s}|a)\in{\cal P}$. \newline
{\it Proof.} As the set $\cal P$ is convex, we have that $\rho_t({\bf s}|a)\equiv
(1-t)\rho({\bf s}|a)+t\rho_1({\bf s}|a)\in {\cal P}$ for every $t\in[0,1]$. As
the function 
$$
{\cal K}_t\equiv \left. {\cal K}\right|_{\rho({\bf s}|a)=\rho_t({\bf s}|a),R=R_0}
$$
 is minimal in $t=1$, we have that
\begin{equation}
\left. \frac{d {\cal K}_t}{d t}\right|_{t=1}\le0,
\end{equation}
which gives Ineq.~(\ref{ineq_lemma2}). $\square$ \newline

\begin{lemma}
\label{lemma_3}
For every pair of distributions $\rho({\bf s}|a)$ and $\rho_1({\bf s}|a)$,
we have that
\begin{equation}
\label{ineq_2_lemma}
\sum_{{\bf s},a}\rho({\bf s},a)\log\frac{\rho({\bf s},a)}{\sum_{\bar a}\rho({\bf s},\bar a)}\ge 
\sum_{{\bf s},a}\rho({\bf s},a)\log\frac{\rho_1({\bf s},a)}{\sum_{\bar a}\rho_1({\bf s},\bar a)},
\end{equation}
where $\rho({\bf s},a)=\rho({\bf s}|a)\rho(a)$ and $\rho_1({\bf s},a)=\rho_1({\bf s}|a)\rho(a)$.
\end{lemma}
{\it Proof.} 
For every differentiable convex function $f(x)$, we have that $f(y)\ge f(x)+(y-x)f'(x)$.
As the function $\sum_{{\bf s},a}\rho({\bf s},a)\log
\frac{\rho({\bf s},a)}{\sum_{\bar a}\rho({\bf s},\bar a)}$
is convex in $\rho({\bf s}|a)$ and its derivative is equal to 
$\rho(a)\log\frac{\rho({\bf s},a)}{\sum_{\bar a}\rho({\bf s},\bar a)}$, we have that
\begin{equation}
\begin{array}{c}
\sum_{{\bf s},a}\rho({\bf s},a)\log\frac{\rho({\bf s},a)}{\sum_{\bar a}\rho({\bf s},\bar a)}\ge 
\sum_{{\bf s},a}\rho_1({\bf s},a)\log\frac{\rho_1({\bf s},a)}{\sum_{\bar a}\rho_1({\bf s},\bar a)}  \\
+\sum_{{\bf s},a}\left[\rho({\bf s},a)-\rho_1({\bf s},a)\right]
\log\frac{\rho_1({\bf s},a)}{\sum_{\bar a}\rho_1({\bf s},\bar a)},
\end{array}
\end{equation}
and, therefore, Ineq.~(\ref{ineq_2_lemma}).
$\square$ \newline

At this point, we can prove the following.\newline
\begin{theorem}
Algorithm~2 converges to the solution of Problem~2.
\end{theorem}
{\it Proof.} 
Let $\rho_n({\bf s}|a)$ and $R_{n-1}({\bf s})$ with $n=1,2,\dots$ be the series 
generated by the algorithm. Namely, $R_n({\bf s})$ is the minimizer of the function 
$\left.{\cal K}\right|_{\rho({\bf s}|a)=\rho_n({\bf s}|a)}$ with respect to
$R({\bf s})$ and $\rho_n({\bf s}|a)$ the minimizer of
$\left.{\cal K}\right|_{R({\bf s})=R_{n-1}({\bf s})}$ with respect to
$\rho({\bf s}|a)$. In other words, the series is generated as follows. We
start with an initial distribution $R_0({\bf s})$ and compute $\rho_1({\bf s}|a)$
through block-minimization of ${\cal K}$ with respect to $\rho({\bf s}|a)$
by taking $R({\bf s})=R_0({\bf s})$. Then, we compute $R_1({\bf s})$ through
block-minimization with respect to $R({\bf s})$ by taking
$\rho({\bf s}|a)=\rho_1({\bf s}|a)$ and so on. The block-minimization 
with respect to $R({\bf s})$ gives 
\begin{equation}
R_n({\bf s})=\sum_a\rho_n({\bf s}|a)\rho(a),
\end{equation}
At the $n$-th round, after the minimization with
respect to $R({\bf s})$, the objective function ${\cal K}$ takes the value
\begin{equation}
{\cal K}_n= 
\sum_{{\bf s},a}\rho_n({\bf s}|a)\rho(a)\log
\frac{\rho_n({\bf s}|a)}{\sum_{\bar a}\rho_n({\bf s}|\bar a)\rho(\bar a)}
\end{equation}
By construction, the series ${\cal K}_n$ is monotonic decreasing. To prove
that the series converges to the minimum ${\cal J}(P)$ of ${\cal K}$, it is sufficient
to prove that
\begin{equation}
\label{liminf_ineq}
\liminf_{n\rightarrow\infty} {\cal K}_n\le{\cal J}(P).
\end{equation}
First, we have that
\begin{equation}
{\cal K}_n\le \sum_{{\bf s},a}\rho_n({\bf s}|a)\rho(a)\log\frac{\rho_n({\bf s}|a)}{R_{n-1}({\bf s})},
\end{equation}
since $R_n({\bf s})$ maximizes $\cal K$ with respect to $R({\bf s})$.
Using Lemma~\ref{lemma_1_ineq}, we obtain the inequalities
\begin{equation}
{\cal K}_n\le \sum_{{\bf s},a}\rho({\bf s}|a)\rho(a)\log\frac{\rho_n({\bf s}|a)}{R_{n-1}({\bf s})}
\end{equation}
for every $\rho({\bf s}|a)\in{\cal V}$.
Thus,
\begin{equation}
\begin{array}{c}
{\cal K}_n\le \sum_{{\bf s},a}\rho({\bf s}|a)\rho(a)\log\frac{\rho_n({\bf s}|a)}{R_n({\bf s})} 
\vspace{1mm} \\
+\sum_{{\bf s},a}\rho({\bf s}|a)\rho(a)\log\frac{R_n({\bf s})}{R_{n-1}({\bf s})}.
\end{array}
\end{equation}
As $R_n({\bf s})=\sum_a\rho_n({\bf s}|a)\rho(a)$, Lemma~\ref{lemma_3} implies that
\begin{equation}
\label{ineq_abc}
\begin{array}{c}
{\cal K}_n\le \sum_{{\bf s},a}\rho({\bf s}|a)\rho(a)\log
\frac{\rho({\bf s}|a)}{\sum_{\bar a}\rho({\bf s}|\bar a)\rho(\bar a)} \\
+\sum_{{\bf s},a}\rho({\bf s}|a)\rho(a)\log\frac{R_n({\bf s})}{R_{n-1}({\bf s})}.
\end{array}
\end{equation}
By making the identifications
\begin{equation}
\begin{array}{c}
\sum_{{\bf s},a}\rho({\bf s}|a)\rho(a)\log
\frac{\rho({\bf s}|a)}{\sum_{\bar a}\rho({\bf s}|\bar a)\rho(\bar a)}\rightarrow c, \\
\sum_{{\bf s},a}\rho({\bf s}|a)\rho(a)\log
\frac{\rho({\bf s}|a)}{R_n({\bf s})}\rightarrow \beta_n,  \\
{\cal K}_n\rightarrow a_n,
\end{array}
\end{equation}
Ineq.~(\ref{ineq_abc}) takes the form of Eq.~(\ref{recur_eq}). 
the quantity $\beta_n$ is not negative for every $n$. Furthermore, it is finite for
$n=0$, since $R_0({\bf s})>0$ (initialization condition in the algorithm). Thus,
Lemma~\ref{lemma_csis} implies that
\begin{equation}
\liminf_{n\rightarrow\infty} {\cal K}_n\le 
\sum_{{\bf s},a}\rho({\bf s}|a)\rho(a)\log
\frac{\rho({\bf s}|a)}{\sum_{\bar a}\rho({\bf s}|\bar a)\rho(\bar a)}
\end{equation}
for every $\rho({\bf s}|a)\in{\cal V}$.
Thus,
\begin{equation}
\liminf_{n\rightarrow\infty} {\cal K}_n\le \min_{\rho({\bf s}|a)\in{\cal V}} I({\bf S};A)={\cal J}(P).
\end{equation}
$\square$

\subsection{Convergence of Algorithm~1}
\label{conv_algo1}
The machinery used for proving the convergence of Algorithm~2 cannot be used
for Algorithm~1, since the distribution $\rho(a)$ is updated at each
round of the iteration. Here, we give some arguments supporting the
hypothesis that also Algorithm~1 converges toward the minimum. This 
hypothesis is also supported by numerical experiments.

As shown in Ref.~\cite{montina2,montina3} and later in Sec.~\ref{sec_nec_suff},
the necessary and sufficient conditions for optimality of Problem~2 are
\begin{equation}
\label{nec_suff_cond}
\begin{array}{c}
\rho({\bf s}|a)=\rho({\bf s}) e^{\sum_b\lambda(s_b,a,b)},   \\
F_\lambda({\bf s})  \le1,  \\
\rho({\bf s})\ge0, \\
\sum_{{\bf s},s_b=s} \rho({\bf s}|a)=P(s|a,b).
\end{array}
\end{equation}
The distributions $\rho({\bf s}|a)$ and $\rho(a)$ are also solutions of 
Problem~1 if $\rho(a)$ maximizes the mutual information $I({\bf S};A)$.

We first observe that the decrease of ${\cal K}$ through the 
block-minimization with respect to $R({\bf s})$ goes to zero as the
number of iterations goes to infinity. That is,
\begin{equation}
\lim_{n\rightarrow\infty} \left[\max_{\rho(a)} {\cal K}_{n-1/2}
-\max_{\rho(a)} {\cal K}_n\right]=0,
\end{equation}
where
\begin{equation}
{\cal K}_{n-1/2}\equiv
\sum_{\bf s,a}\rho_n({\bf s}|a)\rho(a)\log\frac{\rho_n({\bf s}|a)}{R_{n-1}({\bf s})},
\end{equation}
\begin{equation}
{\cal K}_n\equiv
\sum_{\bf s,a}\rho_n({\bf s}|a)\rho(a)\log\frac{\rho_n({\bf s}|a)}{R_n({\bf s})},
\end{equation}
$\rho_n({\bf s}|a)$ and $R_n({\bf s})$ being the distributions 
$\rho({\bf s}|a)$ and $R({\bf s})$ at the $n$-th interation. 
Thus,
\begin{equation}
\limsup_{n\rightarrow\infty} \left[{\cal K}_{n-1/2}-{\cal K}_n\right]_{\rho(a)=\rho_n(a)}\le 0,
\end{equation}
where $\rho_n(a)$ maximizes ${\cal K}_n$. This gives the inequality
\begin{equation}
\limsup_{n\rightarrow\infty} 
\sum_{\bf s}R_n({\bf s})\log\frac{R_n({\bf s})}{R_{n-1}({\bf s})}\le 0.
\end{equation}
The terms of the sequence are the relative entropy between $R_n({\bf s})$ and 
$R_{n-1}({\bf s})$ and are always non-negative. Thus,
\begin{equation}
\lim_{n\rightarrow\infty} 
\sum_{\bf s}R_n({\bf s})\log\frac{R_n({\bf s})}{R_{n-1}({\bf s})}=0.
\end{equation}
Since the relative entropy between two distributions is equal to zero only if
the distributions are equal, we also have
\begin{equation}
\lim_{n\rightarrow\infty}\left[R_n({\bf s})-R_{n-1}({\bf s})\right]=0.
\end{equation}

Now, the minimization at the $n$-th iteration gives
\begin{equation}
\rho_n({\bf s}|a)=R_{n-1}({\bf s})e^{\sum_b\lambda_n(s_b,a,b)},
\end{equation}
We also have $R_{n-1}({\bf s})\simeq R_{n}({\bf s})=\rho_n({\bf s})$ with
arbitrary precision, provided that $n$ is arbitrary large.
Thus,
\begin{equation}
\rho_n({\bf s}|a)\simeq \rho_n({\bf s})e^{\sum_b\lambda_n(s_b,a,b)}
\end{equation}
with arbitrary precision, which is the first optimality 
condition~(\ref{nec_suff_cond}). Also the third and fourth conditions
are satisfied. Thus, it remains to check if the second condition
is asymptotically satisfied in the limit $n\rightarrow\infty$. 
Let us assume that sequences $\rho_n({\bf s}|a)$ and $\rho_n(a)$, as well
as $\lambda_n$, converge to some limit point. In particular, it is
sufficient to assume that $F_{\lambda_n}({\bf s})$ converges to
some $F_\lambda({\bf s})$. Thus, it is clear from step~\ref{update_R} that
$F_\lambda({\bf s})\le 1$, otherwise $R_n({\bf s})$ would explode to infinity
for every $\bf s$ such that $F_\lambda({\bf s})>1$. Note that $R_n({\bf s})$ 
converges to a nonzero value only if $F_\lambda({\bf s})=1$. Indeed,
the first condition for optimality implies that
\begin{equation}
\rho({\bf s})\ne 0 \Rightarrow F_\lambda({\bf s})=1.
\end{equation}
Given for granted that the sequences $\lambda_n$ and $\rho_n(a)$ converge to 
some $\lambda$ and $\rho(a)$, respectively,
this reasoning shows that ${\cal K}_n$ converges toward the minimum of
$\cal K$ with $\rho(a)$ equal to the limit distribution. Furthermore, it 
converges to the minimum of $\bar{\cal K}$, since
$\rho_n(a)$ is the minimizer of the mutual information at each step 
of the iteration. Thus, the algorithm converges to the solution of Problem~1.

\section{Error estimation}
\label{sec_error}

The iterations of Algorithm~1 stop at step~\ref{stop_item} when a given accuracy
is reached. Until now, we have not addressed the issue of how to provide an
estimate of the error.
As the algorithm converges to $C_{min}^{asym}$ from above, the value of 
$I({\bf S};A)$ obtained in each iteration yields an upper bound.
In the following we will use the dual form of problem~2 to derive a lower bound
and we will show that the difference of the bounds converges to zero 
and is thus a reasonable measure of the accuracy of $C_{min}^{asym}$.
We will employ the necessary and sufficient conditions for optimality~(\ref{nec_suff_cond})
derived in Ref.~\cite{montina2} to do so.

The section is organized as follows. In Sec.~\ref{dual_sec}, we introduce the 
dual form of Problem~2, which takes the form of a geometric 
program~\cite{montina2,montina3}. Then, In Sec.~\ref{sec_converg}, we show
how to compute lower and upper bounds at each step of the iteration.
In Sec.~\ref{sec_nec_suff}, we use the dual problem to derive necessary
and sufficient conditions for optimality. Using the conditions, we show
that, in the limit of infinite iterations, the lower and upper bounds 
approach the asymptotic communication complexity.
Thus, as a possible stopping criterion, the iterations are terminated when 
the difference between the lower and upper bound is below some accuracy.

\subsection{Dual form of Problem~2}
\label{dual_sec}
The dual form of a constrained minimization problem (primal problem) is a 
maximization problem in 
which the constraints are replaced by variables, the Lagrange multipliers. 
In general, the dual maximum is smaller than the minimum of the primal problem. 
The difference between the minimum and the maximum is called the duality gap.
However, the dual maximum turns out to be equal to the primal minimum
if some regularity conditions on the constraints of the primal problem
are satisfied~\cite{boyd}. This is the case for Problem~2~\cite{montina2,montina3}.

The dual objective function is given by the minimum of the Lagrangian with respect
to the primal variables. As done for the derivation of the numerical algorithm,
it is convenient to replace the objective function $I(A;{\bf S})$ of Problem~2
with function $\cal K$ defined by Eq.~(\ref{new_obj_funct}). The minimization
is now performed on the variables $\rho({\bf s}|a)$ and $R({\bf s})$. The first
variables satisfy the constraints of Problem~1. Additionally, the variables $R({\bf s})$
satisfy the positivity constraints
\begin{equation}
R({\bf s})\ge 0.
\end{equation}

A direct way for getting the dual problem passes through the dual form
of the block optimization over the variables $\rho({\bf s}|a)$. As we have
seen in Sec.~\ref{min_rho_sa}, this dual form is an unconstrained maximization
of the objective function ${\cal K}_1$, given by Eq.~(\ref{Eq_K_1}).
Thus, Problem~2 takes the form
\begin{equation}
\begin{array}{c}
\min_{R({\bf s})}\max_{\lambda(s,a,b)} {\cal K}_1  
\vspace{1mm} \\
\text{subject to the constraints}  
\vspace{1mm}  \\
R({\bf s})\ge0.
\end{array}
\end{equation}
The variables $R({\bf s})$ in ${\cal K}_1$
can be regarded as Lagrange multipliers associated with
the inequality constraints of the following maximization problem:
\begin{problem}
\begin{equation}
\begin{array}{c}
\max_{\lambda(s,a,b)} {\cal I}_{dual}
\vspace{1mm} \\
\text{subject to the constraints}  
\vspace{1mm}  \\
F_\lambda({\bf s})\le1,
\end{array}
\end{equation}
\end{problem}
where the objective function is 
\begin{equation}
{\cal I}_{dual}\equiv\sum_{s a b}P(s|a,b)\rho(a)\lambda(s,a,b).
\end{equation}
Problem~3 is the dual form of Problem~2 and was derived in Ref.~\cite{montina2,montina3}
in different ways. It is a particular case of geometric program~\cite{gp1,gp2}.

\subsection{Lower and upper bounds on ${\cal C}_{min}^{asym}$}
\label{sec_converg}
A lower bound on the asymptotic communication complexity is provided by any feasible
point of the dual Problem~3. This gives a lower bound on the optimal value of Problem~2
and, hence, a lower bound on the optimal value of Problem~1. A feasible point can be
easily obtained at each step of Algorithm~1. The procedure is as follows. Given the
Lagrange multipliers $\lambda(s,a,b)$ computed at step~2 of the algorithm, we define
the variables $\tilde\lambda(s,a,b)\equiv\lambda(s,a,b)+k$ by adding a constant to
$\lambda(s,a,b)$. The constant is chosen so that $\tilde\lambda(s,a,b)$ satisfy the
constraints of Problem~3, that is, we have
\begin{equation}
\label{const_low_bound}
F_{\tilde\lambda}({\bf s})\le1.
\end{equation}
The quantity 
\begin{equation}
{\cal C}_-=\sum_{s,a,b}P(s|a,b)\rho(a)\left[\lambda(s,a,b)+k\right]
\end{equation}
is a lower
bound on ${\cal C}_{min}^{asym}$. To have a lower bound as close as possible to
${\cal C}_{min}^{asym}$, we have to choose the constant $k$ such that ${\cal C}_-$
is as large as possible and the constraints~(\ref{const_low_bound})
are satisfied. This is attained by taking 
$k=-|B|^{-1}\log\max_{\bf s} F_\lambda({\bf s})$, which gives the lower bound
\begin{equation}
{\cal C}_{-}= \sum_{s,a,b}P(s|a,b)\rho(a)\lambda(s,a,b)-
\log\max_{\bf s} F_\lambda({\bf s}).
\end{equation}

An upper bound on ${\cal C}_{min}^{asym}$ at each step of the iteration is given by 
the value taken by the objective function ${\cal I}_0$. After each iteration, we have
that
\begin{equation}
\label{variables_after_step}
\begin{array}{c}
\rho({\bf s}|a)=\rho({\bf s})F_\lambda^{-1}({\bf s}) e^{\sum_b\lambda(s_b,a,b)}, 
\vspace{1mm}   \\
R({\bf s})=\rho({\bf s}), 
\vspace{1mm}   \\
\sum_{{\bf s},s_b=s}\rho({\bf s}|a)=P(s|a,b).
\end{array}
\end{equation}
Using the first two equations, the upper bound takes the form
\begin{equation}
\label{upper_bound}
{\cal C}_{+}\equiv \sum_{{\bf s},a,b}\rho({\bf s}|a) \rho(a)\lambda(s_b,a,b)-
\sum_{\bf s} \rho({\bf s})\log F_\lambda({\bf s}).
\end{equation}
Using the last of Eqs.~(\ref{variables_after_step}), the upper bound can 
be rewritten in the form
\begin{equation}
\label{upper_bound_2}
{\cal C}_{+}=\sum_{s,a,b}P(s|a,b) \rho(a)\lambda(s,a,b)-
\sum_{\bf s} \rho({\bf s})\log F_\lambda({\bf s}),
\end{equation}
which is computationally more convenient, as the summation over the sequence 
${\bf s}=\{s_1,\dots,s_{|B|}\}$ is replaced by the summation over $s$.

As Algorithm~1 converges to the solution of Problem~1, ${\cal C}_+$ obviously
converges to ${\cal C}_{min}^{asym}$ from above. To prove that the lower bound 
${\cal C}_-$ also converges to ${\cal C}_{min}^{asym}$ from below, we need
the necessary and sufficient conditions for optimality introduced in 
Ref.~\cite{montina2}. This will be done in Sec.~\ref{sec_nec_suff}.

\subsection{Convergence of the lower bound}
\label{sec_nec_suff}
\subsubsection{Necessary and sufficient conditions for optimality}
Let us derive the necessary and sufficient conditions for optimality introduced in 
Ref.~\cite{montina2}.
Every feasible point of the primal Problem~2 and the dual Problem~3 provide upper and lower
bound on ${\cal C}_{min}^{asym}$, respectively. Thus, necessary and sufficient conditions
for optimality are given by the primal and dual constraints and the condition that
the primal and dual objective functions are equal, that is,
\begin{equation}
{\cal I}_{dual}=I(A;{\bf S}).
\end{equation}
This condition is equivalent to the equation
\begin{equation}
\sum_{{\bf s},a}\rho({\bf s}|a)\rho(a)\log\frac{\rho({\bf s}|a)}{\rho({\bf s}) e^{\sum_b\lambda(s_b,a,b)}}=0,
\end{equation}
which can be written in the form
\begin{equation}
\label{ident_cond}
\sum_{{\bf s},a}\rho({\bf s}|a)\rho(a)\log\frac{\rho({\bf s}|a)\rho(a)}{\tilde\rho({\bf s},a)}+ 
\sum_{{\bf s}}\rho({\bf s})\log F_{\lambda}^{-1}({\bf s}) =0,
\end{equation}
where $\tilde\rho({\bf s},a)$ is the probability distribution
\begin{equation}
\tilde\rho({\bf s},a)\equiv\rho({\bf s}) F_\lambda^{-1}({\bf s})  \rho(a)e^{\sum_b\lambda(s_b,a,b)}
\end{equation}

The first term in the left-hand side of Eq.~(\ref{ident_cond}) is the relative entropy between the probability
distributions $\rho({\bf s}|a)\rho(a)$ and $\tilde\rho({\bf s},a)$, 
and it is always non-negative~\cite{cover}. The relative entropy is equal to zero if and only if the
two probability distributions are equal. The dual inequality constraints also imply that the second term is
non-negative. Thus, the equality is satisfied if and only if the two terms are equal to zero, that is, 
if
\begin{equation}
\begin{array}{c}
\rho({\bf s}|a)=\rho({\bf s}) F_\lambda^{-1}({\bf s})e^{\sum_b\lambda(s_b,a,b)} 
\vspace{1mm} \\
\rho({\bf s})\ne 0 \Rightarrow  F_\lambda({\bf s})=1.
\end{array}
\end{equation}
These equations are equivalent to 
\begin{equation}
\rho({\bf s}|a)=\rho({\bf s}) e^{\sum_b\lambda(s_b,a,b)}.
\end{equation}
Thus, Eqs.~(\ref{nec_suff_cond}) are necessary and sufficient conditions for optimality
of Problem~2.

The solution of Problem~1, which gives the asymptotic communication complexity, has 
an extra-condition. The problem is equivalent to the minimax problem defined by 
Eqs.~(\ref{swap_minmax},\ref{eq_J}). As the mutual information is convex in
$\rho({\bf s}|a)$ and concave in $\rho(a)$, the distributions $\rho({\bf s}|a)$
and $\rho(a)$ are solutions of the minimax problem if and only if $\rho({\bf s}|a)$
is a solution of Problem~\ref{prob2} and $\rho(a)$ maximizes the mutual information
$I(A,{\bf S})$. This can be shown by using the minimax theorem.
It is possible to show by using the method of the Lagrange multipliers that
the distribution $\rho(a)$ maximizes the mutual information if and only if
\begin{equation}
\sum_{{\bf s}}\rho({\bf s}|a)\log\frac{\rho({\bf s}|a)}{\rho({\bf s})}\le
\sum_{{\bf s},\bar a}\rho({\bf s}|\bar a)\rho(\bar a)
\log\frac{\rho({\bf s}|\bar a)}{\rho({\bf s})}.
\end{equation}
Using the conditions~(\ref{nec_suff_cond}), this equation can be concisely written
in the form
\begin{equation}
\sum_{s \bar a b} P(s|a,b)\left[\delta_{a,\bar a}-\rho(\bar a)\right]\lambda(s,\bar a,b)\le0.
\end{equation}

\subsubsection{Proof of the lower bound convergence}
As ${\cal C}_+$ converges to ${\cal C}_{min}^{asym}$, to prove that also the lower
bound converges to  ${\cal C}_{min}^{asym}$, it is sufficient
to show that the difference 
\begin{equation}
\Delta{\cal C}\equiv
{\cal C}_+-{\cal C}_-=\log\max_{\bf s}F({\bf s})-\sum_{\bf s}\rho({\bf s})\log F({\bf s})
\end{equation}
goes to zero. The first condition for optimality~(\ref{nec_suff_cond}) and the first of
Eqs.~(\ref{variables_after_step}) imply that $\rho({\bf s})\log F({\bf s})$ goes to $0$
as the algorithm approaches the solution. The second condition for optimality also implies
that $\max_{\bf s} F({\bf s})$ goes to $1$. Thus, $\Delta{\cal C}$ goes to
zero.

In conclusion, as a stopping criterion, we employ the condition
\begin{equation}
\Delta{\cal C}\le \xi,
\end{equation}
where $\xi$ is some given accuracy on the asymptotic communication complexity.
This criterion guarantees that the algorithm will stop and that the error on ${\cal C}_{min}^{asym}$
is smaller than $\xi$. Actually, $\Delta{\cal C}$ can be a very loose estimate of 
the error. Indeed, the actual error ${\cal C}_+-{\cal C}_{min}^{asym}$ generally
scales quadratically with respect to $\Delta{\cal C}$.

\section{Numerical simulations}
\label{num_simu_section}
In Sec.~\ref{sec_num_algo}, we have introduced a simple algorithm for numerically
computing the asymptotic communication complexity ${\cal C}_{min}^{asym}$ by solving
the minimax problem~\ref{prob1}. In this section, we illustrate the method with some
numerical examples. 
In particular, we consider the following scenario. Alice prepares a single qubit and
sends it to Bob who then performs a projective measurement. In this case the two-dimensional quantum
state can be represented by a three-dimensional Bloch vector. 
So Alice prepares one of $|A|$ possible quantum states characterized by its Bloch vector $\vec v_a$, 
with $a\in \left\{1,\ldots,|A|\right\}$. After receiving $\vec v_a$ through the \emph{noiseless} quantum channel,
Bob performs one of $|B|$ projective measurements on the qubit.
The projective measurement is completely characterized by the
eigenstates associated with the measurement outcomes, in this case a pair of opposite normalized Bloch vectors, 
say $\pm\vec w_b$ with $b\in \left\{1,\ldots, |B|\right\}$. 
Let us associate $\pm\vec w_b$ with the outcome values $s=\pm 1$.
Thus, the conditional probability $P(s|a,b)$ takes the form
\begin{equation}
P(s|a,b)=\frac{1}{2}\left(1+s\; \vec v_a\cdot \vec w_b\right).
\end{equation}
First, we consider the case of Bloch vectors $\vec v_a$ and $\vec w_b$ being equidistributed on a plane.
Then the analytical solution of Problem~\ref{prob1} is known~\cite{montina}. 
Namely, we take
\begin{equation}
\label{states_meas}
\vec v_a=\left(
\begin{array}{c}
\cos\frac{2\pi a}{|A|} \vspace{1mm} \\
\sin\frac{2\pi a}{|A|} \vspace{1mm} \\
0
\end{array}
\right),  \;\;\;
\vec w_b=\left(
\begin{array}{c}
\cos\frac{\pi b}{|B|} \vspace{1mm} \\
\sin\frac{\pi b}{|B|} \vspace{1mm} \\
0
\end{array}
\right).
\end{equation}
Note that the vectors $\vec w_b$ cover only a half-plane, as the opposite vectors correspond
to the same measurements with the outcomes interchanged. 

\begin{figure}
\epsfig{figure=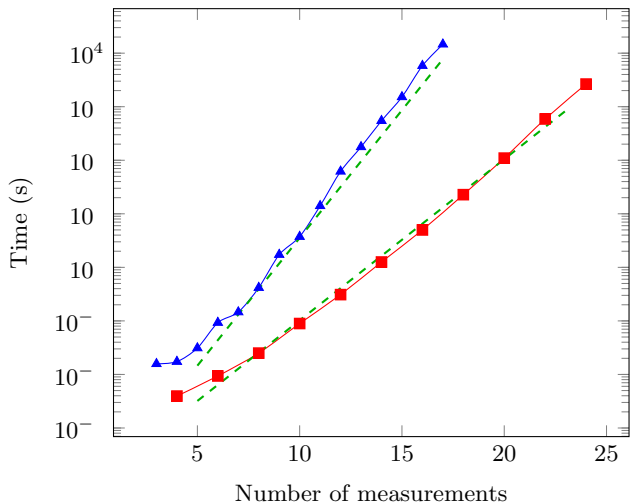}
\caption{Computational time as a function of the number of measurements for noiseless
quantum channel with capacity $1$ qubit. The measurements and states are in planar
configuration on the Bloch sphere. Data from Mosek library are represented by triangles
and data from Algorithm~2 by squares. They are compared with the functions
$6\times 10^{-5}3^{|B|}$ and $10^{-4}2^{|B|}$, respectively
(dashed line).}
\label{fig_times}
\end{figure}

Since the conditional probability is invariant under the transformation
$a\rightarrow a+1$ and $b\rightarrow b+1$ up to a swap of $s$, the uniform distribution 
$\rho(a)=1/|A|$ is a solution of the maximization in Eq.~(\ref{swap_minmax}).
Indeed, suppose that $\rho_m(a)$ is a solution. By symmetry, also $\rho_m(a+k)$ is
solution for every constant integer $k$. Since the objective function ${\cal J}(P)$ is
concave, also the uniform distribution $|A|^{-1}\sum_{k=1}^{|A|}\rho_m(a+k)$ is a solution.
As $\rho(a)$ is known, the computation of ${\cal C}_{min}^{asym}$ is performed
through Algorithm~2. In Fig.~\ref{fig_times}, we show the corresponding
computational time as a function of the number of measurements $|B|$ (red line
with squares). The blue line with triangles represents the computational time
of the Mosek package. The accuracy is $10^{-6}$. For a large number of measurements,
the computational time of Algorithm~1 and Mosek grows roughly as  $2^{|B|}$ and 
$3^{|B|}$, respectively, as shown by the green dashed lines.

%

\begin{figure}
\epsfig{figure=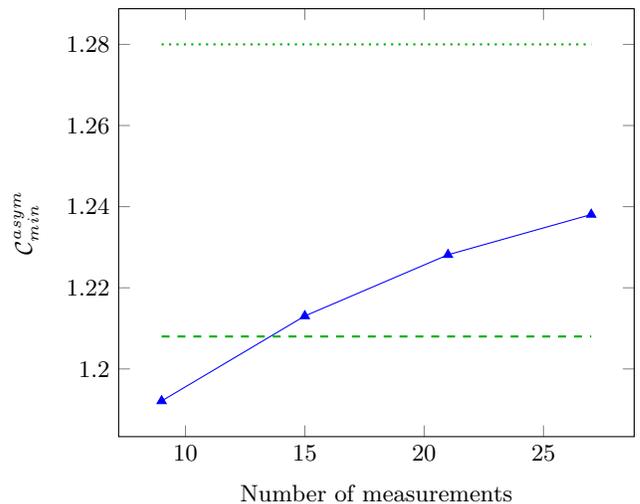}
\caption{Asymptotic communication complexity for the non-planar configuration with
$9$, $15$, $21$, and $27$ measurements. The dashed line is the previous lower bound
1.208 bits, evaluated with infinite planar states and measurements~\cite{montina}. 
The dotted line is the upper bound 1.28 bits, derived in Ref.~\cite{montina4}.}
\label{fig2}
\end{figure}

In Ref.~\cite{montina}, we found that the asymptotic communication complexity
approaches the value $1+\log_2(\pi/e)\simeq 1.208$ in the limit of infinitely many 
planar states and measurements. Thus, this value provides a lower bound on
the asymptotic communication complexity of a noiseless quantum channel
with capacity $1$ qubit with infinite states and measurements densely covering
the Bloch sphere. This lower bound can be improved by considering a 
nonplanar configuration. Namely, in addition to the vectors~(\ref{states_meas})
in the plane $x-y$, we add similar vectors in the planes $x-z$ and $y-z$, for both, Alice and Bob. Let
$A_0$ be the number of equidistributed states in each plane. If $A_0$ is a multiple of $4$,
the vectors along the axes $x$, $y$, and $z$ are shared by two planes. Thus,
the overall number of states is $3(A_0-2)=|A|$. Similarly, if the number
of measurements, say $B_0$, is even, then the measurements with eigenvectors
along the coordinate axes are shared by two planes. The overall number
of measurements is $3(B_0-1)=|B|$. Let us take $A_0=2 B_0$, so that the
set of states is equal to the overall set of the eigenvectors associated with
the measurements. We have considered the cases $B_0=4,6,8$, and $10$, which 
correspond to $9$, $15$, $21$, and $27$ overall measurements, respectively.
The value $B_0=12$ corresponds to $33$ measurements, which would require 
more than $200$~GB of memory and too long computational time with our 
available hardware. The obtained asymptotic communication complexity
is depicted in Fig.~\ref{fig2}. Whenever $|B|>9$, the obtained
values are larger than the previous lower bound $1.208$ bits, indicated by the green dashed line.
In particular, we improve the best lower bound to $1.238$ bits with $27$ measurements.

\section{Conclusion}

The computation of the communication complexity of a quantum communication process
can be reduced to the computation of the minimal capacity over a suitable set
of classical channels. The advantage of this reduction is provided by the convexity
of the optimization problem, implying that every local minimum is global. However,
the capacity of channels does not have in general an analytic form, but it is given
as a maximum over the input distribution. Thus, the optimization problem takes
the form of a minimax problem and cannot be solved directly by using optimization
libraries for convex optimization. 

In this paper, we have presented a numerical method that solves the minimax problem.
To compare the performance of the method with the commercial Mosek package, we
have performed numerical experiments for quantum processes whose associated 
objective function takes an analytical form. Compared to Mosek, our method
turns out to be significantly faster and displays a better scaling law with respect to
the number of states and measurements.
As a further illustration of the method, we have improved the previously known
lower bound $1.208$ on the asymptotic communication complexity of a noiseless quantum
channel with capacity $1$ qubit. Nonetheless, there remains a significant gap between the new 
computed lower bound, $1.238$ bits, and the known upper bound $1.28$~\cite{montina4}. 
Thus, the question which value is optimal remains open.
There are however reasons to believe that the upper bound is the optimal value. 
It has been derived with an explicit protocol for infinite states and projective
measurements. The adaptation of that protocol to the planar case gives the correct 
optimal value of $1.208$ bits derived in Ref.~\cite{montina}.

In spite of its simplicity, the introduced algorithm displays very good performance.
A slight change in the update step~\ref{update_R} in Algorithm~1b can further 
improve the convergence properties. Namely, this step can be replaced by
\begin{equation}
R({\bf s})\rightarrow R({\bf s})\left[F_\lambda({\bf s})\right]^\alpha,
\end{equation}
where $\alpha$ is some number greater than $1$, appropriately chosen for accelerating
the convergence. In a subsequent paper, we will introduce a more sophisticated
algorithm for computing the minimal communication cost of general communication
complexity problems.

{\it Acknowledgments.} 
This work is supported by the Swiss National Science Foundation, the NCCR QSIT, 
the COST action on Fundamental Problems in Quantum Physics and Hasler foundation
through the project number 14030 "Information-Theoretic Analysis of Experimental 
Qudit Correlations".

\bibliography{biblio.bib}

\end{document}